\documentclass[conference, 10pt]{IEEEtran}

\usepackage{amsmath,amssymb,amsfonts}
\usepackage{algorithmic}
\usepackage{graphicx}
\usepackage{textcomp}
\usepackage{xcolor}
\usepackage{multirow}

\usepackage{enumitem}
\usepackage{booktabs}
\usepackage{pifont}
\usepackage{color}
\usepackage{tikz}
\usepackage{multirow}
\usepackage{float}
\usepackage{listings}
\usepackage{soul}
\usepackage{fontawesome}
\usepackage{textcomp}  
\usepackage{scalerel}  

\usepackage[hidelinks]{hyperref}
\usepackage{makecell}

\newcommand*\circled[1]{\raisebox{.4pt}{\tikz[baseline=(char.base)]{\node[shape=circle,draw,inner sep=1pt, style={fill=black, text=white}, scale=0.75] (char) {\textbf{#1}};}}}

\definecolor{codegreen}{rgb}{0,0.6,0}
\definecolor{codegray}{rgb}{0.5,0.5,0.5}
\definecolor{codepurple}{rgb}{0.58,0,0.82}
\definecolor{backcolour}{rgb}{0.99,0.99,0.99}

\lstdefinestyle{codestyle}{
    backgroundcolor=\color{backcolour},   
    commentstyle=\color{codegreen},
    keywordstyle=\color{magenta},
    numberstyle=\tiny\color{codegray},
    stringstyle=\color{codepurple},
    basicstyle=\linespread{0.8}\ttfamily\small,
    breakatwhitespace=false,         
    breaklines=true,                 
    captionpos=b,                    
    keepspaces=true,                 
    numbers=left,                    
    numbersep=5pt,                  
    showspaces=false,                
    showstringspaces=false,
    showtabs=false,                  
    tabsize=1,
    escapechar=\%,
    columns=fullflexible,
    frame=single,
    xleftmargin=3.4pt,
    xrightmargin=3.4pt,
}

\makeatletter
\newcommand\fs@plainbold{\def\@fs@cfont{\color{blue}\bfseries}\let\@fs@capt\floatc@plain
\def\@fs@pre{}\def\@fs@post{}%
\def\@fs@mid{\vspace\abovecaptionskip\relax}%
\let\@fs@iftopcapt\iffalse}
\makeatother
\floatstyle{plainbold}
\newfloat{lstfloat}{htbp}{lop}
\floatname{lstfloat}{Listing}

\newcommand{\ourwork}{{GNNBuilder}}

\begin{document}

\title{\fontsize{24}{24}\selectfont GNNBuilder: An Automated Framework for Generic Graph Neural Network Accelerator Generation, Simulation, and Optimization}

\DeclareRobustCommand{\IEEEauthorrefmark}[1]{\smash{\textsuperscript{\footnotesize #1}}}

\author{
\IEEEauthorblockN{Stefan Abi-Karam\IEEEauthorrefmark{1}\IEEEauthorrefmark{2}, Cong Hao\IEEEauthorrefmark{1}}
\IEEEauthorblockA{\IEEEauthorrefmark{1}School of Electrical and Computer Engineering, Georgia Institute of Technology\\
\IEEEauthorrefmark{2}Georgia Tech Research Institute\\
\{stefanabikaram, callie.hao\}@gatech.edu}
}

\maketitle

\thispagestyle{plain}
\pagestyle{plain}

\begin{abstract}
There are plenty of graph neural network (GNN) accelerators being proposed. However, they highly rely on users' hardware expertise and are usually optimized for one specific GNN model, making them challenging for practical use . Therefore, in this work, we propose GNNBuilder, the first automated, generic, end-to-end GNN accelerator generation framework. It features four advantages: (1) GNNBuilder can automatically generate GNN accelerators for a wide range of GNN models arbitrarily defined by users; (2) GNNBuilder takes standard PyTorch programming interface, introducing zero overhead for algorithm developers; (3) GNNBuilder supports end-to-end code generation, simulation, accelerator optimization, and hardware deployment, realizing a push-button fashion for GNN accelerator design; (4) GNNBuilder is equipped with accurate performance models of its generated accelerator, enabling fast and flexible design space exploration (DSE). In the experiments, first, we show that our accelerator performance model has errors within $36\%$ for latency prediction and $18\%$ for BRAM count prediction. Second, we show that our generated accelerators can outperform CPU by $6.33\times$ and GPU by $6.87\times$. This framework is open-source, and the code is available at \url{https://github.com/sharc-lab/gnn-builder}.
\end{abstract}

\section{Introduction}

Graph Neural Networks (GNNs) are a powerful and popular tool for solving learning tasks where the data can be represented as a graph. Among different applications, GNNs can be used for node-level, edge-level, and graph-level tasks, such as drug discovery \cite{moleculenet}, recommender systems \cite{wu_graph_2022}, social network analysis \cite{benamira_semi}, traffic forecasting \cite{eta_traffic}, electronic health records analysis \cite{golmaei_deepnote}, scene graph understanding \cite{chang_comprehensive_2022}, electronic design automation \cite{wu_ironman_pro_2022}, natural language processing \cite{wu_graph_2021}, autonomous driving \cite{pointacc}, and high-energy physics \cite{elabd_graph_2022}.
Among these applications, some have real-time constraints for GNN inference and require hardware acceleration. One example is autonomous driving systems that use GNNs to process LIDAR point cloud data \cite{li_deep_2020}. Another prominent example is in high-energy physics, where GNNs are used for real-time particle detection~\cite{elabd2022graph} and jet lag detection~\cite{qu2020jet}, which must be processed within several nano-second.

Given the acceleration needs for GNN inference, there are many GNN accelerators being proposed. Examples include earliest ASIC accelerators proposed by Auten et al.~\cite{auten2020hardware}, HyGCN~\cite{hygcn}, and EnGN~\cite{engn}, as well as most recent accelerators such as AWB-GCN~\cite{awbgcn}, BoostGCN~\cite{zhang2021boostgcn}, I-GCN~\cite{igcn}, GCNAX \cite{gcnax}, Rubik~\cite{rubik}, and GraphACT~\cite{graphact}.
Among them, Rubik and GraphACT aim to accelerate GCN training using ASIC and FPGA, respectively.

Despite the great success of GNN accelerators, there are still significant \textbf{limitations}. \underline{First}, \textit{existing GNN accelerators are model-specific but not generic}. Specifically, most GNN accelerators focus on only one or two most popular GNN models, such as Graph Convolution Network (GCN)~\cite{kipf2016semi} or GraphSage~\cite{hamilton2017inductive}, and provide fixed accelerator structures, fixed GNN layer types, activations, and other design choices that are specific to the implemented model.
These accelerators are not generic and \textit{cannot handle advanced GNNs such as anisotropic GNNs, GNNs with edge embeddings, or complicated aggregation functions}~\cite{tailor2021we, xu2018powerful, corso2020principal}. The fundamental reason is that most existing GNN accelerators simplify GNN computations to be a sequence of general or sparse matrix multiplications, which \textit{does not hold true} for those advanced GNNs.
\underline{Second}, \textit{most of the accelerators are hard-coded and require extensive hardware expertise to adapt to new GNN models}. There is no existing tools that can generate GNN accelerators automatically, optimally, and without any hardware knowledge.
There are only two existing works that can support automated accelerator generation: DeepBuring-GL \cite{deepburninggl} and HP-GNN \cite{lin2022hp}. DeepBuring-GL targets inference acceleration but is limited to a fixed GCN or GraphSAGE model. HP-GNN targets training acceleration but not real-time-inference. Moreover, HP-GNN proposes its own model API and lacks the flexibility to support a wide range of GNN architectures and different features. Table~\ref{tab:compare-with-existing} summarizes the limitations of DeepBuring-GL and HP-GNN. Therefore, researchers and practitioners cannot explore the best GNN model for their target applications in software and easily deploy their application-specific models to hardware for acceleration.

Motivated by the existing limitations of GNN accelerator designs and tools, we propose {\ourwork}, a generic, feature-rich, and extensible framework for end-to-end GNN accelerator generation, simulation, optimization, and deployment on FPGAs with bitstreams. To be generic, we follow the \textit{message passing mechanism} of GNN models, which can express almost all types of GNN models at the theoretical formulation level, as stated by a recent work~\cite{velickovic_2022}. To be extensible, we directly take \textit{standard PyTorch} as the programming language, which allows programmers to design their own GNN models freely and can be directly used for training.

\begin{itemize}[leftmargin=*]

 \item {\textbf{Generic: wide range of GNN model support}. Our proposed framework, {\ourwork}, offers a wide range of support for GNN models through an explicit message passing approach. In addition to supporting state-of-the-art models such as GCN, GIN, GraphSAGE, and PNA, {\ourwork} allows for the customization of various features such as layer type, activation, quantization, aggregation, and pooling. This level of customization is not offered in HP-GNN, as summarized in Table~\ref{tab:compare-with-existing}.}
 
\item {\textbf{Extensibility: Interoperability with PyTorch}. {\ourwork} is the first work that allows users to define their model architectures freely in native PyTorch using a parameterizable GNNModel PyTorch module. This allows users to seamlessly integrate accelerator design as part of existing deep learning workflows. Therefore, {\ourwork} not only supports standard GNNs (as listed in Table~\ref{tab:GNN-list}) but can extend to almost all customized GNN models supported in PyTorch Geometric.}
 
\item{\textbf{Support for node-level and graph-level tasks + node and edge input features}. Our \textit{\ourwork} supports node-level and graph-level task outputs, as well as node-level and edge-level feature inputs. This allows \ourwork to support a wide range of acceleration applications, including drug screening, high-energy physics, and point cloud processing.}

\item{\textbf{Accelerator Design Space Exploration (DSE) and Optimization}. \textit{\ourwork} provides tools to help designers automatically select the best configurations for the generated accelerator, such as hardware parallelism, resource allocation, and fixed-point precision, instead of manual exploration. This automated DSE can significantly improve performance in seconds, as opposed to days, to achieve the best latency under fixed resource constraints with a trade-off in model accuracy.}

\item {\textbf{Open-source Python API with end-to-end workflow}. Our {\ourwork} provides an open-source Python library with APIs that allow users to define their own models from development to deployment in a push-button fasion with zero hardware expertise required. It is an end-to-end workflow including: \circled{a} hardware-compatible simulation, \circled{b} testbench build and execution, \circled{c} automated hardware code generation and synthesis, and deployment on FPGA with host code.}

\item{\textbf{Superior performance against CPU and GPU.} {\ourwork} generates high-performance accelerators on FPGA that outperform PyTorch Geometric CPU and GPU baselines on various datasets by  \textbf{6.33$\times$} and \textbf{6.87$\times$} respectively.}

\end{itemize}



\begin{table}[h]
    \scriptsize
    \centering    
    \caption{GNNBuilder Comparison with Existing Work}
    \setlength\tabcolsep{3pt}
    
    \begin{tabular}{c | c | c |c }
    \toprule
    
    & \textbf{HP-GNN~\cite{lin2022hp}} & \textbf{DeepBurning-GL~\cite{deepburninggl}} & \textbf{\ourwork} \\
    \midrule
    
    Acceleration Goal & Training & Inference & Inference \\ \hline
    
    Programming Language & Self-defined & PyTorch and DGL & PyTorch \faThumbsUp \\
    
    Anisotropic GNN Family & No & No & Yes \faThumbsUp \\ 
    
    Extensibility & Low & Low & Very High \faThumbsUp \\ \hline
    
    Arbitrary Quantization & No & No & Yes \faThumbsOUp \\
    
    Arbitrary Aggregation & No & No & Yes \faThumbsOUp \\

    Arbitrary Activation & Fixed & Fixed & Arbitrary \faThumbsOUp \\
    
    Skip Connections & No & No & Yes \faThumbsOUp \\
    
    Arbitrary Global Pooling & No & No & Yes \faThumbsOUp \\ 
    
    Arbitrary MLP Head & No & No & Yes \faThumbsOUp \\
    
    \makecell{Fixed + Floating Point\\Testbench} & No & No & Yes \faThumbsOUp \\ 
    
    Open Source & No & No & Yes \faThumbsOUp \\
    
    \bottomrule
    \end{tabular}
    
    \label{tab:compare-with-existing}
\end{table}

\begin{table}
    \centering
    \small
    
    \caption{Supported GNNs types by our framework \ourwork~(also supports user-defined GNN models)
    }

    \begin{tabular}{p{0.11\textwidth}| p{0.3\textwidth}}
    \toprule
    \textbf{Model} & \textbf{Representativeness} \\
    \midrule
    \textbf{GCN} \cite{kipf2016semi} & GNN family that can be represented as sparse matrix-matrix multiplications (SpMM) \\ \hline
    
    \textbf{GraphSAGE} \cite{hamilton2017inductive} & GNN family with flexible / non-sum aggregation methods \\ \hline
    
    \textbf{GIN} \cite{xu2018powerful} & GNN family with \textit{edge embeddings}, SpMM \textit{does not} apply  \\ \hline

    \textbf{PNA} \cite{corso2020principal} & A popular Anisotropic GNN family arbitrarily using multiple aggregation methods and sophisticated message function, SpMM \textit{does not} apply\\ \hline

    \multicolumn{2}{p{0.45\textwidth}}{\textbf{GCN:} graph convolutional network; \textbf{GIN:} graph isomorphism network; \textbf{GraphSAGE:} graph sample and aggregate; \textbf{PNA:} principal neighborhood aggregation.} \\
    
    \bottomrule
    
    \end{tabular}
    
    \label{tab:GNN-list}
    
\end{table}

 \section{Related Work and Motivations}
 \label{sec:related}

\subsection{Related Work}

\subsubsection{GNN Accelerators and Graph Accelerators}

The increasing use of Graph Neural Networks (GNNs) in real-time and large-data applications in the research community and industry has led to numerous GNN accelerator studies. A recent survey \cite{abadal2021computing} provides an overview of GNN accelerators for CPU, GPU, ASIC, FPGA, and heterogeneous platforms. Some specific GNN accelerators include Auten et al. \cite{auten}, HyGCN \cite{hygcn}, AWB-GCN \cite{awbgcn}, EnGN \cite{engn}, GRIP \cite{grip}, GCNAX \cite{gcnax}, Rubik \cite{rubik}, GraphACT \cite{graphact}, Boost-GCN \cite{zhang2021boostgcn}, and I-GCN \cite{igcn}. These accelerators explore different implementations and model-specific design choices to achieve speedups in GNN inference and training. More recent accelerators, such as GenGNN \cite{abi2022gengnn} and FlowGNN \cite{flowgnn}, also adopt a GNN model agnostic approach for inference acceleration without sacrificing performance.

\subsubsection{GNN Accelerator Automation}

Some existing works explore the automated generation of hardware accelerators for GNNs.  One key work is DeepBuring-GL \cite{deepburninggl} which is focused on generating GNN inference accelerators for CPU-FPGAs systems such as Xilinx's Alveo U50. However this work only support a fixed subset of GCN-based architectures. Another work, HP-GNN \cite{lin2022hp}, also targets acceleration but for GNN training on CPU-FPGA platforms. HP-GNN also supports a subset GCN-based and GraphSAGE-based architectures. 

\subsection{Limitations}

\subsubsection{GNN Accelerators}

Existing GNN acceleration approaches primarily focus on fixed model architectures for inference and often support only isotropic models, which allows them to leverage sparse matrix multiplication acceleration techniques. These approaches generally implement GCN or GIN architectures by simplifying computations with sparse matrix multiplications (SpMM) and general matrix multiplications (GEMM). However, advanced GNNs cannot be reduced to mere matrix multiplications and require specialized graph preprocessing and model computation patterns that are not easily generalizable to anisotropic models.

The limitations of these approaches stem from their focus on optimization techniques that hinder generalization to more advanced GNN architectures. In contrast, recent works such as GenGNN and FlowGNN propose hardware architectures that can accommodate advanced model architectures with anisotropic message passing support by adopting an explicit message passing hardware dataflow. This offers a more flexible solution for a broader range of GNN models.

\subsubsection{GNN Accelerator Automation}

Current accelerator automation approaches, as shown in Table \ref{tab:compare-with-existing}, have limitations in generalizing to advanced GNN architectures. DeepBurning-GL and HP-GNN allow end-to-end code generation but are limited to GCN and GraphSAGE models. They lack support for anisotropic GNNs like PNA, expressive GNNs such as GIN, and features like mean and variance neighbor pooling, arbitrary activation functions, skip connections, sum/mean/max global pooling, and MLP prediction heads. Additionally, they do not offer simple fixed-point quantization or code generation for fixed-point and floating-point testbenches, essential for rapid debugging. These limitations restrict the applicability of existing frameworks for researchers and practitioners working with diverse GNN models.

\section{{\ourwork} Framework Overview}
\label{sec:overview}

\begin{figure*}[h]
    \centering
    \includegraphics[width=1.0\textwidth]{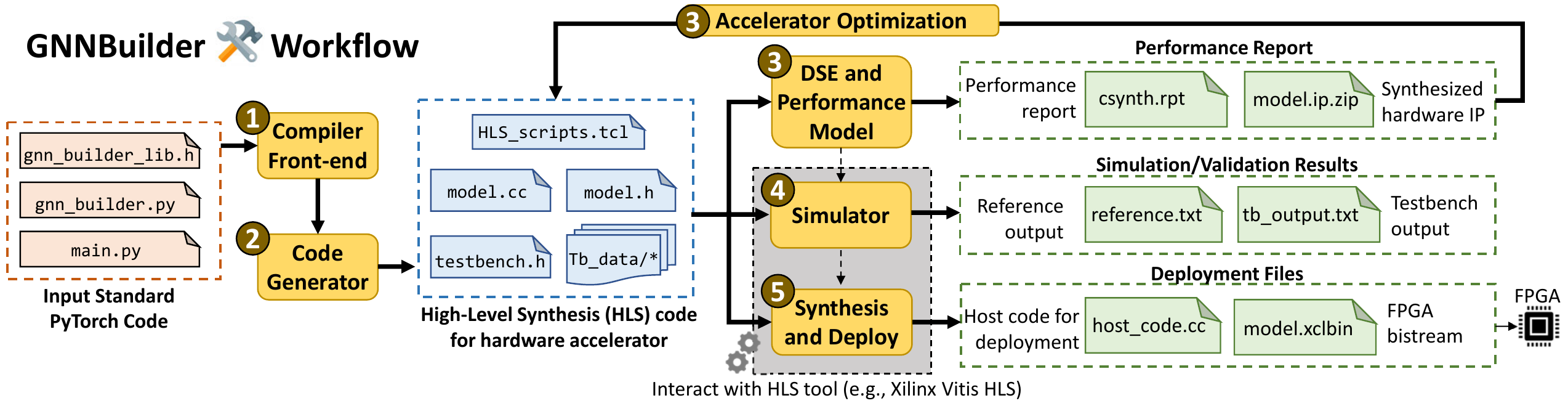}
    \caption{Workflow of the GNNBuilder framework. }
    \label{fig:overview}
\end{figure*}

\subsection{{\ourwork} Components}

{\ourwork} aims to provide users with a streamlined process to design, implement, validate, and optimize GNN models, transitioning from standard PyTorch models to FPGA bitstreams. As depicted in Fig.~\ref{fig:overview}, {\ourwork} consists of five components:
\circled{1} \textbf{Compiler front-end} parses the native PyTorch GNN model definition, including the number of GNN layers, layer type, activation type, data precision, pooling type, aggregation type, and MLP definition.
\circled{2} \textbf{Code generator} builds upon a library of pre-defined hardware accelerator templates that adopt the message passing mechanism, making it compatible with various GNN types. We generate High-Level Synthesis (HLS) code targeting FPGAs, supported by the Xilinx Vitis HLS tool \cite{VitisHLS}.
\circled{3} \textbf{Design space exploration and performance model} enables automated DSE for accelerator generation, encompassing hardware parallelism, resource allocation, and quantization (data precision).
\circled{4} \textbf{Simulation and testbench} facilitates transparent hardware-compatible simulation using automatically generated testbenches, ensuring the correctness of accelerator functionality. It also generates plain C++ code for "true" quantization simulation, accurately reflecting on-FPGA quantization accuracy.
\circled{5} \textbf{Hardware synthesis and deployment} automatically generates hardware synthesis scripts, synthesizes FPGA bitstreams, and produces host code for executing the bitstream. Table~\ref{tab:GNN-list} presents the representative GNNs supported by our framework. Although these are examples, {\ourwork} can flexibly accommodate a wide range of customized GNN models, including residual and skip connections, arbitrary quantization, aggregation functions, graph attention, activation, global pooling, and MLP head. Such user-defined features can be naturally expressed using PyTorch, granting {\ourwork} exceptional extensibility.

\begin{table}[]
    \centering
    \small
    \caption{User APIs for the \texttt{gnn\_builder} Python library.}
    \resizebox{\columnwidth}{!}{
    \begin{tabular}{l|c}
    \toprule
    \textbf{API Functions}     & \textbf{Description}  \\ \midrule
    \texttt{model.GNNModel(nn.Module)} & PyTorch Model for GNNBuilder Arch. \\
    \texttt{model.GCNConv\_GNNB(nn.Module)} & GCN Conv. Layer \\
    \texttt{model.GINConv\_GNNB(nn.Module)} & GIN Conv. Layer \\
    \texttt{model.PNAConv\_GNNB(nn.Module)} & PNA Conv. Layer \\
    \texttt{model.SAGEConv\_GNNB(nn.Module)} & GraphSAGE Conv. Layer \\
    \texttt{model.GlobalPooling(nn.Module)} & Global Graph Pooling Layer \\
    \texttt{model.MLP(nn.Module)} & MLP Prediction Head \\
    \midrule
    \texttt{code\_gen.Project()}     &  GNNBuilder Project Class \\
    \texttt{Project.gen\_hw\_model()} & Code Gen. For HW Kernel \\
    \texttt{Project.gen\_testbench()} & Code Gen. For Testbench \\
    \texttt{Project.gen\_makefile()} & Code Gen. For Testbench Makefile \\
    \texttt{Project.gen\_vitis\_hls\_tcl\_script()} & Code Gen. For Vitis HLS Synth. Script \\
    \texttt{Project.build\_and\_run\_testbench()} & Build and Run Testbench \\
    \texttt{Project.run\_vitis\_hls\_synthesis()} & Launch Vitis HLS Synthesis Run \\
    
    \bottomrule
    \end{tabular}
    }
    
    \label{tab:gnn-api}
\end{table}

\subsection{Programming Model and User APIs}

Table~\ref{tab:gnn-api} showcases the user APIs provided by {\ourwork}, and Listing 1 illustrates an example of the user interface for a customized GNN model.

To begin, a user defines a \texttt{GNNModel} instance, incorporating an \texttt{MLP} and a \texttt{xxxConv\_GNNB} module (e.g., \texttt{PNAConv\_GNNB}). {\ourwork} offers wrapper classes for each graph convolution layer, enabling the user to specify parallelism factors \texttt{p\_in} and \texttt{p\_out}. The higher-level \texttt{GNNModel} supports arguments for defining architecture parameters and separate parallelism factors for the GNN head (\texttt{gnn\_p\_in}, \texttt{gnn\_p\_hidden}, \texttt{gnn\_p\_out}) and the MLP head (\texttt{p\_in}, \texttt{p\_hidden}, \texttt{p\_out}). The user can train and manipulate the \texttt{GNNModel} instance as a standard PyTorch module.

A user can then define a GNNBuilder \texttt{Project} instance. The \texttt{Project} class has several arguments to define build paths, the \texttt{GNNModel} model instance, the PyTorch Geometric dataset for the model task, \texttt{max\_nodes} and \texttt{max\_edges}, numerical precision, and average number of nodes, edges, and node in-degree for synthesis runtime estimation.

After creating a \texttt{Project} instance, the user can call the code generation functions to produce the model kernel HLS code, the kernel testbench code and data, the testbench makefile, and the Vitis HLS build script. Post code generation, the user can call \texttt{build\_and\_run\_testbench()} to build and execute the testbench, and \texttt{run\_vitis\_hls\_synthesis()} to execute the Vitis HLS synthesis process. These execution scripts also return data for the testbench runtime, mean absolute error (MAE), and synthesis latency / resource usage.

\begin{lstlisting}[
    language=python,
    frame=single,
    numbers=none,
    captionpos=b,
    caption={Example usage of GNNBuilder Framework.},
    label={lst:user-api}
  ]
import torch.nn as nn
from torch_geometric.datasets import MoleculeNet

import gnnbuilder as gnnb

dataset = MoleculeNet(root="./tmp/MoleculeNet", name="hiv")

model = gnnb.GNNModel(
    graph_input_feature_dim=dataset.num_features,
    graph_input_edge_dim=dataset.num_edge_features,
    gnn_hidden_dim=16,
    gnn_num_layers=2,
    gnn_output_dim=8,
    gnn_conv=gnnb.SAGEConv_GNNB,
    gnn_activation=nn.ReLU,
    gnn_skip_connection=True,
    global_pooling=gnnb.GlobalPooling(["add", "mean", "max"]),
    mlp_head=gnnb.MLP(in_dim=8 * 3, out_dim=dataset.num_classes, hidden_dim=8, hidden_layers=3, activation=nn.ReLU, p_in=8, p_hidden=4, p_out=1),
    output_activation=None,
    gnn_p_in=1,
    gnn_p_hidden=8,
    gnn_p_out=4
)

MAX_NODES = 600
MAX_EDGES = 600
num_nodes_avg, num_edges_avg = gnnb.compute_average_nodes_and_edges(dataset)
degree_avg =  gnnb.utils.compute_average_degree(dataset)

proj = gnnb.Project(
    "gnn_model",
    model,
    "classification_integer",
    VITIS_HLS_PATH,
    BUILD_DIR,
    dataset=dataset,
    max_nodes=MAX_NODES,
    max_edges=MAX_EDGES,
    num_nodes_guess=num_nodes_avg,
    num_edges_guess=num_edges_avg,
    degree_guess=degree_avg,
    float_or_fixed="fixed",
    fpx=FPX(32, 16),
    fpga_part="xcu280-fsvh2892-2L-e",
    n_jobs=32,
)

proj.gen_hw_model()
proj.gen_testbench()
proj.gen_makefile()
proj.gen_vitis_hls_tcl_script()
proj.gen_makefile_vitis()

tb_data = proj.build_and_run_testbench()
print(tb_data)
synth_data = proj.run_vitis_hls_synthesis()
print(synth_data)
\end{lstlisting}

\section{GNNBuilder Model Architecture}
\label{sec:GNNBuilder-arch}

\begin{figure*}[h]
    \centering
    \vstretch{1.0}{\includegraphics[width=0.95\textwidth]{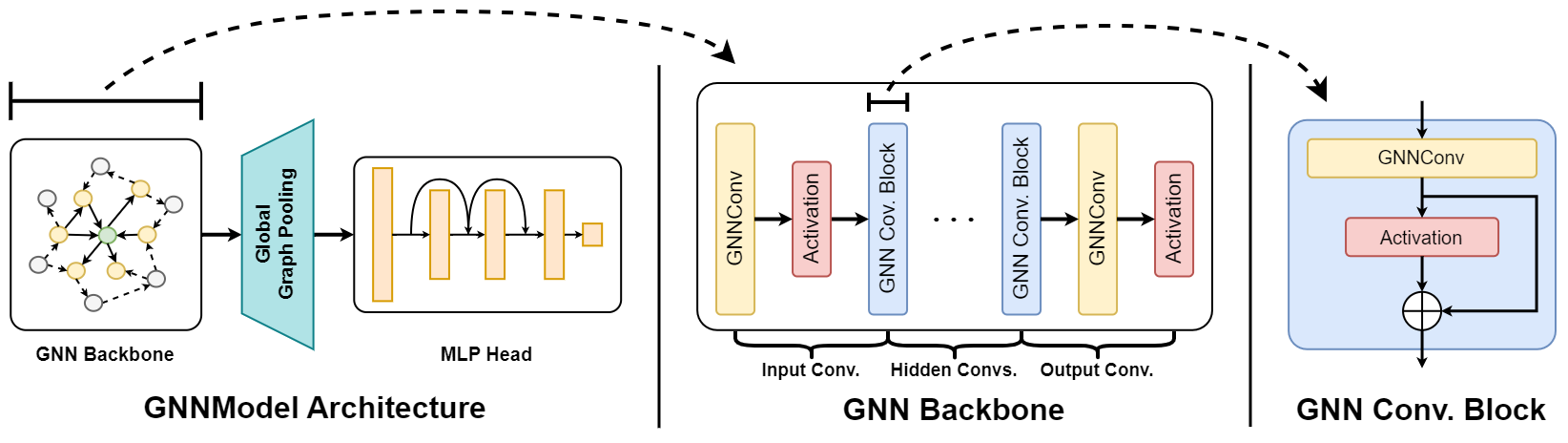}}
    \caption{The GNNBuilder model architecture for graph level tasks.}
    \label{fig:gnnbuilder_arch}
\end{figure*}

Each GNN model in \texttt{gnn\_builder} framework is based on a parameterized \texttt{GNNModel} (subclass of \texttt{torch.nn.Module}) architecture, designed to work seamlessly within the PyTorch ecosystem.

\ourwork~supports node-, edge-, and graph-level tasks using a simple linear model architecture (Fig.~\ref{fig:gnnbuilder_arch}). The \textbf{GNN Backbone} consists of graph convolution layers, activations, and skip connections, with customizable parameters. Supported GNNConv layers include GCN, GraphSAGE, GIN, and PNA. For edge and node tasks, users can remove the pooling and MLP head. The \textbf{Global Graph Pooling} module aggregates node embeddings using sum, mean, or max pooling. The \textbf{MLP Head} transforms the pooled output for the specified task, with customizable input/output embedding sizes, hidden layers, and activation functions. 

These models are defined using existing PyTorch and PyTorch Geometric layers, with user-provided keyword arguments for customization. The template-based compiler matches components from a \texttt{GNNModel} class and parameters to code templates in the HLS code generation output.

\section{Accelerator Architecture}
\label{sec:hardware}

Our accelerator implementation adopts an explicit message-passing architecture that implements dataflow optimization within the GNN Backbone, individual GNN Conv. Layers, and the MLP head. This maximizes latency while implementing efficient streaming data movement using FIFO streams rather than memory buffers. This is the main optimization that shows the best performance gains.

\begin{figure*}[h]
    \centering
    \small
    \includegraphics[width=0.85\textwidth]{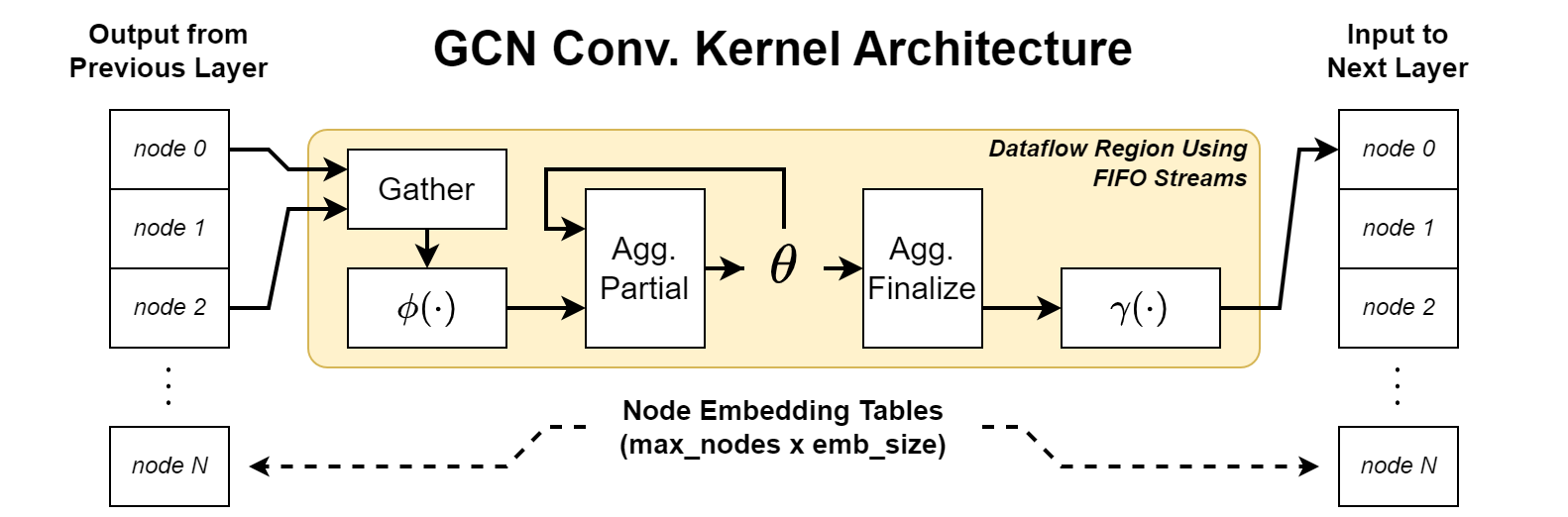}
    \caption{The high-level hardware kernel architecture for GNNConv layers.}
    \label{fig:gnnconv}
\end{figure*}

\subsection{Message Passing and Graph Convolution Kernels}

Inspired by GenGNN \cite{abi2022gengnn} and FlowGNN \cite{flowgnn}, we adopt an explicit message-passing architecture for graph convolution layers, allowing us to support GNN layers like PNA, which are not compatible with traditional SpMM accelerator approaches.

For each node, the operations illustrated in Figure \ref{fig:gnnconv} are performed. We first gather the node's neighbor indices using the neighbor table and offset table. Then, we iterate through each neighbor index to load its associated embedding from the input node embedding table, transform the embedding with $\phi(\cdot)$, and aggregate it with a partial aggregation. After processing all neighbors, we finalize the partial aggregation, combine it with the current node embedding, and transform it with the apply function $\gamma(\cdot)$. The resulting computed embedding is then written to the output node embedding table.

The functions $\phi(\cdot)$, $\gamma(\cdot)$, and the aggregation(s) used depend on the specific layer being implemented. Kernels for GCN, GraphSAGE, GIN, and PNA layers are included in the initial GNNBuilder kernel library.

Developing custom kernels is possible by contributing the appropriate hardware kernel code for the layer of interest to \ourwork's kernel template library, as well as providing a matching GNNConv class that links the kernel in \ourwork's Python library. This can be accomplished through a pull request with minimal effort, and the rest of our framework remains agnostic to the specific types a user intends to add support for.

\subsection{Other Components}

\textit{Graph Data}: In the model kernel, buffers depend on the number of nodes ($\text{num\_nodes}$) or edges ($\text{num\_edges}$), with buffer sizes set to an upper bound determined by $\text{MAX\_NODES}$ and $\text{MAX\_EDGES}$ parameters in a \ourwork~Project instance.
Model kernels require input graphs in \textbf{COO}rdinate format matrix with an input node feature table. The COO matrix is a $\text{MAX\_EDGES} \times 2$ integer array, while the input node feature table is a $\text{MAX\_NODES} \times \text{input\_dim}$ fixed-point datatype array. In-degree and out-degree buffers have a size of $\text{MAX\_NODES}$.
Additionally, a neighbor table stores each node's neighbors, and a neighbor offset table indexes each node's block of neighbors, both sized $\text{MAX\_EDGES}$ and $\text{MAX\_NODES}$, respectively.

\textit{Degree + Neighbor Table Computation}: Before model computation, the degree table of the input graph must be calculated. Node degrees are used by various graph convolutions for normalization purposes. Since these values are only known at runtime, the in-degree and out-degree tables need to be computed on-the-fly in the accelerator for each input graph. The COO format of input graphs allows for computation within the bounds of $\texttt{num\_edges}$. Subsequently, the neighbor table and neighbor offset table are computed simultaneously using two loops: one iterating over $\texttt{num\_edges}$ and the other over $\texttt{num\_nodes}$.

\textit{Partial Aggregations}: To efficiently aggregate neighbor embeddings with constant memory ($O(1)$ space complexity), \ourwork~defines single-pass algorithms for aggregation that avoid the need for buffering all neighbor embeddings in an intermediate buffer, which would consume substantial BRAMs. \ourwork~supports sum, min, max, mean, variance, and standard deviation aggregations. Each aggregation is associated with a data structure for storing partial and final aggregation data. For variance, Welford's one-pass algorithm \cite{var_welford} is used to compute variance efficiently. 

\textit{Linear Layer}: \ourwork~ implements tiled matrix multiplication for linear layers, enabling hardware parallelization. The parallelization factor for each linear layer is controlled by the $\text{BLOCK\_SIZE\_IN}$ and $\text{BLOCK\_SIZE\_OUT}$ template arguments for the linear kernel function. These arguments determine the partition factors for input, weight, and bias arrays, thus controlling the parallelism of the multiply-accumulate (MAC) operations.

\textit{Global Pooling}: \ourwork~supports sum, mean, and max global graph pooling, aggregating node embeddings across all nodes into a single embedding of the same size. Multiple pooling methods can be combined using concatenation.

\textit{Activations}: \ourwork~supports ReLU, Sigmoid, Tanh, and GELU \cite{gelu} activations, implemented using fixed-point math functions from the Vitis HLS fixed-point math library.

\section{Accelerator Generation and Implementation}
\label{sec:generation}

Automated kernel generation is a key advantage of \ourwork, allowing seamless conversion of software models defined using PyTorch into hardware accelerators. This approach reduces development friction by eliminating the need for customized APIs. \ourwork~efficiently generates code through dynamic introspection of software model objects, combined with a template-based compiler and a pre-defined kernel library.

\subsection{Kernel Code Generation}

\ourwork is built on a template-based compiler which facilitates code generation by generating C++ HLS code for the top-level model kernel and associated header directly from a PyTorch model. This enables conditional and loop control flows for template blocks, useful for features like skip-connections, double-buffer array selection, and mapping layer kernel calls in the correct order with accurate input/output size.

The parameterized structure of the GNNModel allows \ourwork~to match appropriate function calls to corresponding kernels from the C++ header-only template library. This approach is extensible, enabling users to add support for other layers, aggregations, or activations by creating associated kernels in the template library and updating the Jinja template.

\subsection{Hardware Simulation and Verification Testbenches}

\ourwork~allows designers to generate and build C++ testbenches for their models, facilitating rapid testing of fixed-point quantizations without synthesizing designs. The testbench code, model parameters, dataset graphs, true output, and PyTorch model outputs are exported as binary files. During runtime, the testbench reads these files, loads weights into the model kernel, evaluates the kernel on all dataset inputs, and compares the output to the PyTorch model outputs.

The testbench calculates verification metrics, such as mean absolute error between the PyTorch-generated model output and the kernel output, and averaged kernel runtime. These values are written to text files during runtime.

For fixed-point models, the testbench ensures accurate fixed-point representations of input graphs and model parameters. By utilizing the Vitis HLS fixed-point library \cite{VitisHLS}, functional equivalence with hardware modeling is maintained. The floating-point data from PyTorch is cast to the user-specified fixed-point format in the testbench.

\subsection{Hardware Deployment on FPGA}
Using the \texttt{run\_vitis\_hls\_synthesis()} function, users can build a synthesized accelerator (Verilog RTL code) and execute the implementation flow to generate Vivado IP blocks (\texttt{.zip}) or Vitis Kernels (\texttt{.xo}) for hardware deployment. This streamlines the workflow from software model to fully implemented design.

\ourwork~supports implementing Vitis kernels on platforms like Alveo U50 and Alveo U280, including full bitstream generation (\texttt{.xclbin}) and a host code testbench for on-chip graph dataset evaluation. This testbench, similar to the C++ testbench, uses Xilinx's runtime library, XRT, and OpenCL for FPGA interfacing from the host CPU.

\section{Performance Model and Design Space Exploration}
\label{sec:dse}

\subsection{FPGA Model Implementation}
\label{ssec:hw-model-impl}

We implemented our models on the Xilinx Alveo U280 FPGA accelerator at 300 MHz using Vitis HLS \cite{VitisHLS} and Vitis \cite{Vitis} tools from Xilinx. Our framework directly provides the generated HLS code to the synthesis tools, accompanied by suitable build scripts.

\subsection{Hardware Performance Model}

To assess the effectiveness of runtime modeling in DSE, we examine direct-fit models for latency and BRAM prediction, comparing them to Vitis HLS-reported post-synthesis values. We focus on BRAM usage for resource modeling, as it is the primary constraint-violating resource.

The direct-fit latency and BRAM models are random forest regressors, fitted on datasets of model configurations and their post-synthesis values. Empirical testing showed that random forests outperformed linear/polynomial models, support vector machines, and gradient boosting tree models in avoiding overfitting.

These direct-fit models necessitate a pre-synthesized design database. As the number of possible configurations is too large for brute-force exploration, sparsely sampling the design space enables fitted models to interpolate between sampled designs, providing accurate estimates for unseen configurations.

\subsection{Design Space Exploration}

Instead of requiring users to run HLS builds for each design configuration or create datasets, we provide serialized trained versions of the direct-fit models described earlier making it feasible to performance brute force or random sampling of the model configuration space. Evaluating these direct-fit models takes milliseconds, compared to minutes for HLS synthesis. This can reduce performance prediction runtime enables users to develop intelligent co-design tools for real-time optimization, paving the way for train-time model sparsity, quantization, and neural architecture search, among other possibilities.

\section{Experimental Setup}
\label{sec:experiments-setup}

\subsection{Hardware Performance Model}

The accuracy of the runtime and BRAM models is assessed against a database of 400 synthesized designs, randomly sampled from a configuration space of model parameters (see Listing \ref{lst:design_sapce}). For the fitted models, a random forest regressor with 10 estimators is used. The models are evaluated using the mean absolute percent error (MAPE) between the true post-synthesis metrics and predicted metrics. To examine overfitting, a 5-fold cross-validation (CV) is conducted, averaging the test MAPE for each fold to obtain the final cross-validation MAPE.

\begin{lstlisting}[
    language=python,
    frame=single,
    numbers=none,
    captionpos=b,
    caption={Design Space used for Hardware Performance Model Dataset},
    label={lst:design_sapce}
  ]
QM9_DATASET = QM9(root="./tmp/QM9").index_select(list(range(1000)))
DATASET_IN_DIM = QM9_DATASET.num_features
DATASET_OUT_DIM = QM9_DATASET[0].y.ravel().shape[0]

MEDIAN_NODES, MEDIAN_EDGES = compute_median_nodes_and_edges(QM9_DATASET, round_val=True)
MEDIAN_DEGREE = compute_median_degree(QM9_DATASET)

MAX_NODES = 600
MAX_EDGES = 600

CONVS = ["gcn", "gin", "pna", "sage"]
GNN_HIDDEN_DIM = [64, 128, 256]
GNN_OUT_DIM = [64, 128, 256]
GNN_NUM_LAYERS = [1, 2, 3, 4]
GNN_SKIP_CONNECTIONS = [True, False]
MLP_HIDDEN_DIM = [64, 128, 256]
MLP_NUM_LAYERS = [1, 2, 3, 4]

GNN_P_HIDDEN = [2, 4, 8]
GNN_P_OUT = [2, 4, 8]
MLP_P_IN = [2, 4, 8]
MLP_P_HIDDEN = [2, 4, 8]
\end{lstlisting}

\subsection{Accelerator Performance Evaluation}
\label{ssec:perf-eval}

This section evaluates various model architecture configurations across multiple datasets, comparing the proposed hardware implementations:

\begin{itemize}
    \item \textbf{PyG-CPU}: A PyTorch Geometric CPU model
    \item \textbf{PyG-GPU}: A PyTorch Geometric GPU model
    \item \textbf{CPP-CPU}: A C++ floating-point CPU model
    \item \textbf{FPGA-Base}: Proposed hardware model without hardware parallelism
    \item \textbf{FPGA-Parallel}: Proposed hardware model with hardware parallelism
\end{itemize}

Performance is analyzed using a fixed GNN model with varying GNNConv layers (GCN, GraphSAGE, GIN, and PNA), across graph-level task datasets such as QM9, ESOL, FreeSolv, Lipophilicity, and HIV from MoleculeNet \cite{moleculenet}.

The CPU models are evaluated on an Intel Xeon Gold 6226R, while the GPU models are assessed on an NVIDIA RTX A6000. The hardware models (FPGA-Base and FPGA-Parallel) are implemented as described in Section \ref{ssec:hw-model-impl}. 

Each baseline is evaluated on a batch size of 1, with the runtimes for CPU and GPU implementations computed by averaging the runtime of the first 1000 graphs of each dataset (or the complete dataset if it contains fewer than 1000 graphs). FPGA implementations' runtimes are obtained from the worst-case estimate provided by Vitis HLS after synthesis. The architecture configuration in Listing \ref{lst:runtime_model_config} is used for all models.

The FPGA-Parallel implementations employ different parallelism factors for GCN, SAGE, and GIN models (\texttt{gnn\_p\_in=1}, \texttt{gnn\_p\_hidden=16}, \texttt{gnn\_p\_out=8}, \texttt{p\_in=8}, \texttt{p\_hidden=8}, \texttt{p\_out=1}), while PNA models use \texttt{gnn\_p\_hidden=8} and \texttt{gnn\_p\_out=8}. These models utilize \texttt{<16, 10>} bit fixed-point data representations. FPGA-Base implementations have parallel factors set to $1$ and implement node features using \texttt{<32, 16>} bit fixed-point types.

\begin{lstlisting}[
    language=python,
    frame=single,
    numbers=none,
    captionpos=b,
    caption={Model Arch. for Benchmark},
    label={lst:runtime_model_config}
]
model = gnnb.GNNModel(
    graph_input_feature_dim=dim_in,
    graph_input_edge_dim=0,
    gnn_hidden_dim=128,
    gnn_num_layers=6,
    gnn_output_dim=64,
    gnn_conv=conv,
    gnn_activation=nn.ReLU,
    gnn_skip_connection=True,
    global_pooling=gnnb.GlobalPooling(["add", "mean", "max"]),
    mlp_head=MLP(in_dim=64 * 3, out_dim=dim_out, hidden_dim=64, hidden_layers=4, activation=nn.ReLU),
    output_activation=None,
)
\end{lstlisting}

\section{Results}
\label{sec:experiments-results}

\subsection{Analytical Performance Model}

\begin{figure}[h]
    \centering
    \includegraphics[width=0.5\textwidth]{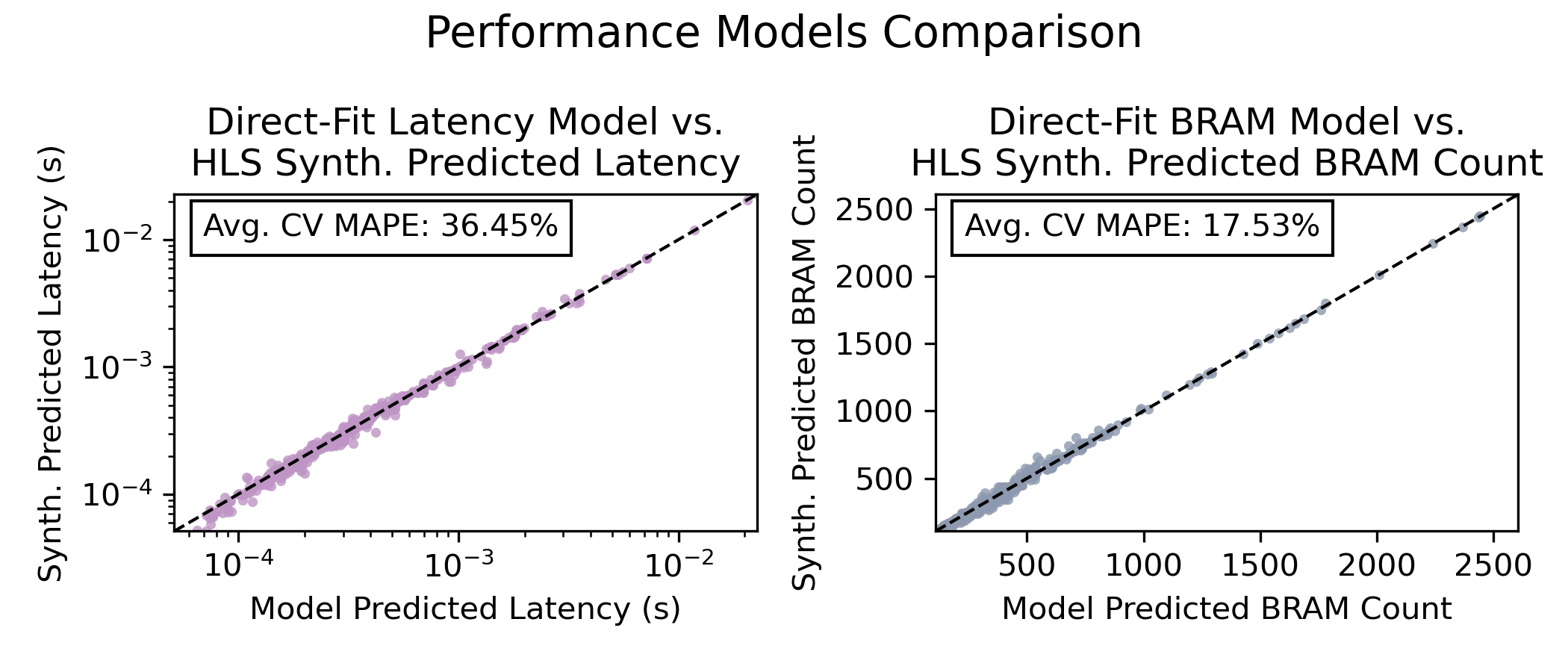}
    \caption{Comparison of latency prediction models with true post-synthesis latency and BRAM usage reported from Vitis HLS}
    \label{fig:latency}
\end{figure}

The results of fitting the latency and BRAM models on our database of generated designs are illustrated in Figure \ref{fig:latency}. The direct-fit latency model achieved a CV MAPE of approximately 36\%, while the direct-fit BRAM model obtained a CV MAPE of approximately 17\%. Figure \ref{fig:latency} demonstrates that the direct-fit model consistently predicts the true value with few outliers. These findings indicate that directly fitting models on a design database, which sparsely samples the design configuration space, is an effective and straightforward approach for performance modeling in \ourwork, enabling rapid Design Space Exploration (DSE).

\subsection{DSE Exploration}

\begin{figure}[h]
    \centering
    \includegraphics[width=0.5\textwidth]{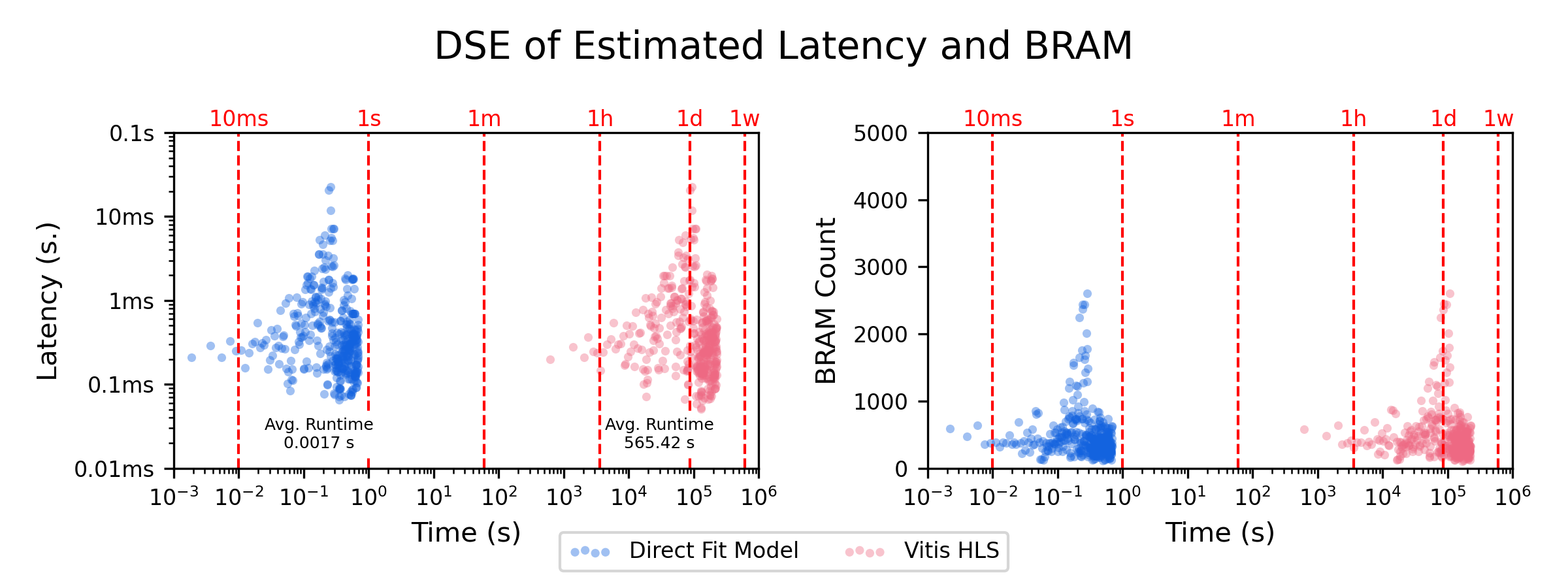}
    \caption{Cumulative runtime for evaluating 400 design to predict model latency and BRAM usage. The x-axis represents time going forward from left to right, and each point represents a performance estimate which has finished computing.} 
    \label{fig:dse}
\end{figure}

To exemplify the speed up of direct fit models over standard evaluation of the HLS tool, we also analyze the performance estimate compute time for all 400 model configurations used to train the direct fit models. We present the results in Figure \ref{fig:dse}, which can be viewed as a timeline of runs. All model calls for the direct fit models to finish in under a second, while all Vitis HLS synthesis runs finish in under two days. An average direct fit model call takes $1.7$ ms, while an average Vitis HLS synthesis run takes $9.4$ minutes. This difference is around 6 orders of magnitude emphasizing the the real-time performance estimation of direct fit models. 

\subsection{Accelerator Performance Evaluation}

\begin{table}[h]
\centering
\small
\caption{FPGA-Parallel speedup over PyG CPU, PyG GPU, and C++ CPU runtimes.}
\label{tab:speedup}
\begin{tabular}{c | ccc}
\toprule
         &  \textbf{PyG-CPU} &  \textbf{PyG-GPU} &  \textbf{CPP-CPU} \\
\midrule
GCN     &                          6.46x &                          7.66x &                          3.04x \\
GIN     &                          5.81x &                          6.08x &                          4.24x \\
PNA     &                          6.48x &                          6.70x &                         22.14x \\
SAGE    &                          6.58x &                          7.16x &                          8.84x \\
Geo. Mean &                          6.33x &                          6.87x &                          7.08x \\ \midrule
\textbf{Geo. Mean} &                          \textbf{6.33x} &                          \textbf{6.87x} &                          \textbf{7.08x} \\

\bottomrule
\end{tabular}
\end{table}

\begin{figure}[h]
    \centering
    \includegraphics[width=0.5\textwidth]{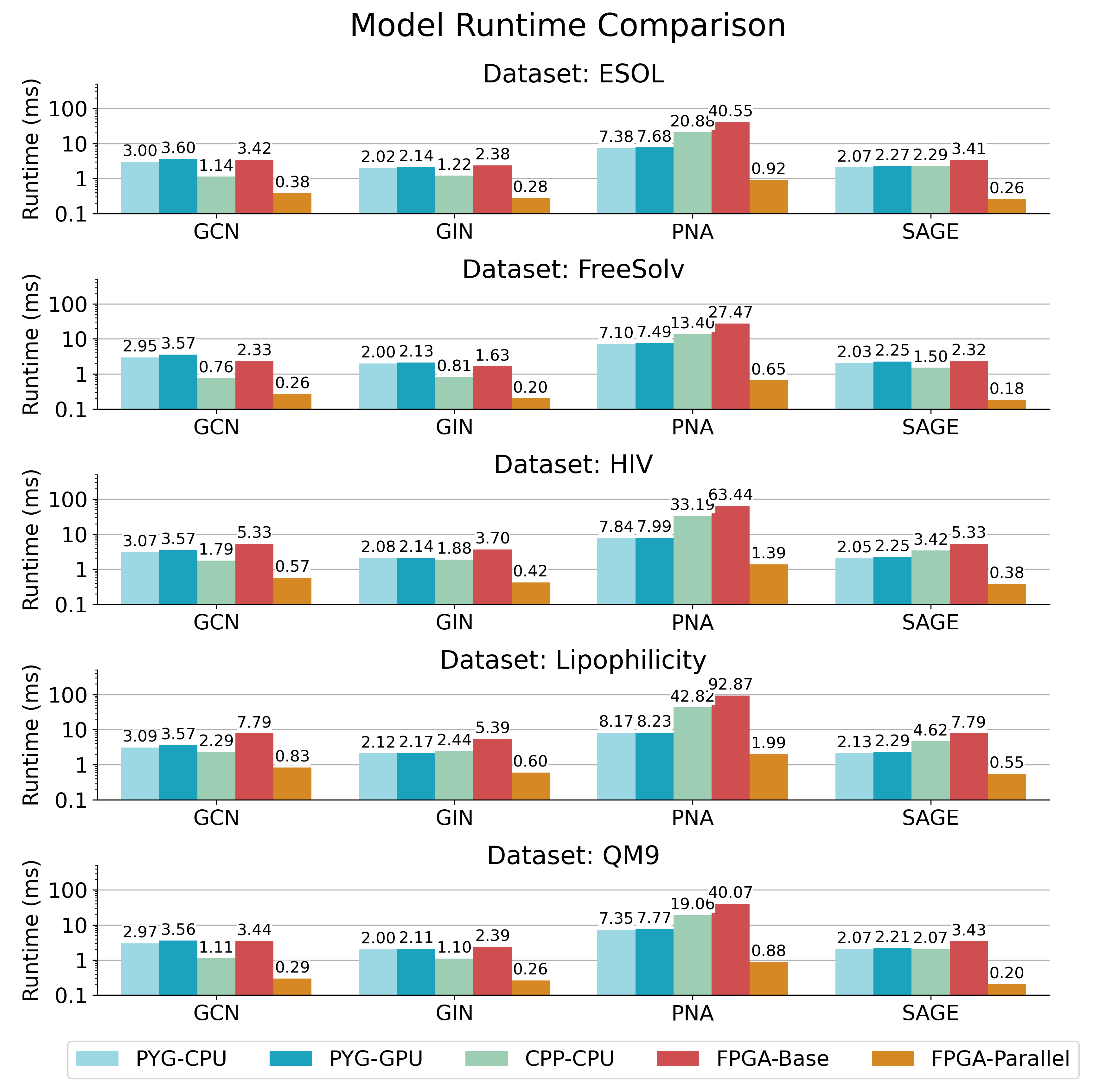}
    \caption{GNN model runtime across a range of architectures, datasets, and implementations (y-axis in log-scale).}
    \label{fig:runtime}
\end{figure}

\begin{figure}[h]
    \centering
    \includegraphics[width=0.5\textwidth]{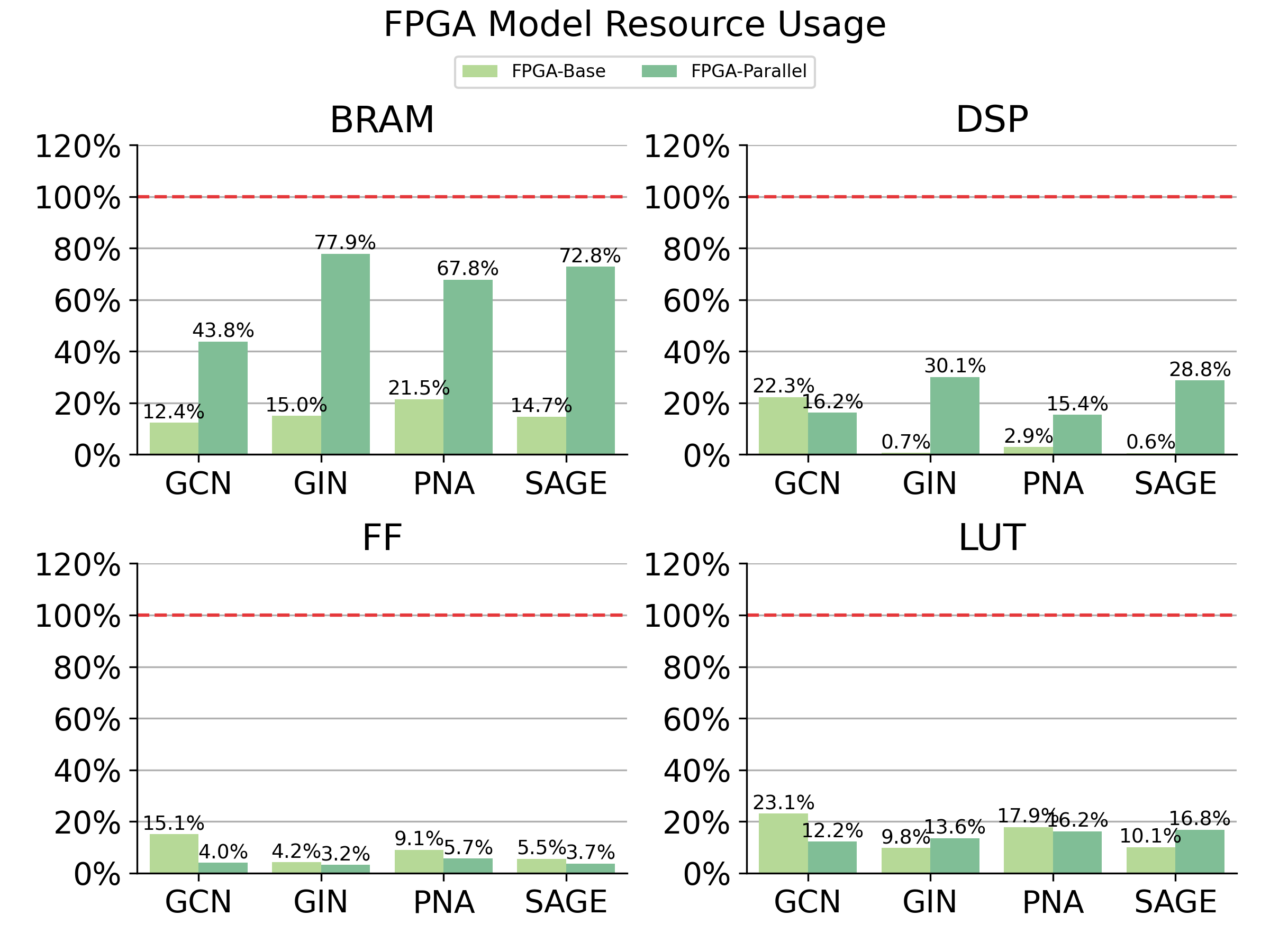}
    \caption{Resource usage of FPGA-Base and FPGA-Parallel model implementations.}
    \label{fig:resources}
\end{figure}

The performance results for the proposed accelerator hardware framework in comparison to other implementations are shown in Figure \ref{fig:runtime}, Figure \ref{fig:resources}, and Table \ref{tab:speedup}. The values in Table \ref{tab:speedup} indicate the speedup factors of teh FPGA-Parallel implementation for the latency values averaged across datasets. For all cases, there is at least a 6x speedup in the parallelized FPGA implementation over the PyG CPU, PyG GPU, and C++ CPU implementations. Across all models, there is a geometric mean speedup of \textbf{6.33$\times$} over PyG-CPU and \textbf{6.87$\times$} over PyG-GPU. The resource usage also shows more room for BRAM and DSP utilization across models indicating there is more room for increased parallelism and higher speedups.

\section{Conclusion}
\label{sec:conclusion}

In this paper, we introduced \textbf{\ourwork}, a versatile end-to-end GNN accelerator generation framework with a user-friendly Python API. \ourwork~supports a wide range of expressive GNNs, seamlessly integrates with PyTorch modules, and offers unique features uncommon in other inference accelerators. We demonstrated its capabilities in generating hardware kernels, testbenches, running testbenches on PyTorch Geometric datasets, and launching Vitis HLS synthesis kernels. Our framework also enables efficient DSE and outperforms CPU and GPU implementations by exploiting hardware parallelism.

The current software framework is available for both software and hardware practitioners at the repo linked below:
\begin{gather*}
\text{\texttt{\url{https://github.com/sharc-lab/gnn-builder}}}
\end{gather*}

Future work involves optimizing graph convolution kernels, exploring intelligent DSE search methods, train-time co-design, and expanding our kernel template library to accommodate more graph convolution kernels such as GAT \cite{velivckovic2017graph} and other emerging GNN architectures.

\bibliography{refs}

\begin{thebibliography}{10}

\bibitem{moleculenet}
Z.~Wu, B.~Ramsundar, E.~N. Feinberg, J.~Gomes, C.~Geniesse, A.~S. Pappu,
  K.~Leswing, and V.~Pande, ``{{MoleculeNet}}: A benchmark for molecular
  machine learning,'' {\em Chemical Science}, vol.~9, no.~2, pp.~513--530,
  2018.

\bibitem{wu_graph_2022}
S.~Wu, F.~Sun, W.~Zhang, X.~Xie, and B.~Cui, ``Graph {Neural} {Networks} in
  {Recommender} {Systems}: {A} {Survey},'' Apr. 2022.
\newblock arXiv:2011.02260 [cs].

\bibitem{benamira_semi}
A.~Benamira, B.~Devillers, E.~Lesot, A.~K. Ray, M.~Saadi, and F.~D. Malliaros,
  ``Semi-{Supervised} {Learning} and {Graph} {Neural} {Networks} for {Fake}
  {News} {Detection},'' in {\em 2019 {IEEE}/{ACM} {International} {Conference}
  on {Advances} in {Social} {Networks} {Analysis} and {Mining} ({ASONAM})},
  pp.~568--569, Aug. 2019.
\newblock ISSN: 2473-991X.

\bibitem{eta_traffic}
A.~Derrow-Pinion, J.~She, D.~Wong, O.~Lange, T.~Hester, L.~Perez, M.~Nunkesser,
  S.~Lee, X.~Guo, B.~Wiltshire, P.~W. Battaglia, V.~Gupta, A.~Li, Z.~Xu,
  A.~Sanchez-Gonzalez, Y.~Li, and P.~Veličković, ``{ETA} {Prediction} with
  {Graph} {Neural} {Networks} in {Google} {Maps},'' in {\em Proceedings of the
  30th {ACM} {International} {Conference} on {Information} \& {Knowledge}
  {Management}}, pp.~3767--3776, Oct. 2021.
\newblock arXiv:2108.11482 [cs].

\bibitem{golmaei_deepnote}
S.~N. Golmaei and X.~Luo, ``{DeepNote}-{GNN}: predicting hospital readmission
  using clinical notes and patient network,'' in {\em Proceedings of the 12th
  {ACM} {Conference} on {Bioinformatics}, {Computational} {Biology}, and
  {Health} {Informatics}}, {BCB} '21, (New York, NY, USA), pp.~1--9,
  Association for Computing Machinery, Aug. 2021.

\bibitem{chang_comprehensive_2022}
X.~Chang, P.~Ren, P.~Xu, Z.~Li, X.~Chen, and A.~Hauptmann, ``A {Comprehensive}
  {Survey} of {Scene} {Graphs}: {Generation} and {Application},'' Jan. 2022.
\newblock arXiv:2104.01111 [cs].

\bibitem{wu_ironman_pro_2022}
N.~Wu, Y.~Xie, and C.~Hao, ``{IronMan}-{Pro}: {Multi}-objective {Design}
  {Space} {Exploration} in {HLS} via {Reinforcement} {Learning} and {Graph}
  {Neural} {Network} based {Modeling},'' {\em IEEE Transactions on
  Computer-Aided Design of Integrated Circuits and Systems}, pp.~1--1, 2022.
\newblock Conference Name: IEEE Transactions on Computer-Aided Design of
  Integrated Circuits and Systems.

\bibitem{wu_graph_2021}
L.~Wu, Y.~Chen, K.~Shen, X.~Guo, H.~Gao, S.~Li, J.~Pei, and B.~Long, ``Graph
  {Neural} {Networks} for {Natural} {Language} {Processing}: {A} {Survey},''
  June 2021.
\newblock arXiv:2106.06090 [cs].

\bibitem{pointacc}
Y.~Lin, Z.~Zhang, H.~Tang, H.~Wang, and S.~Han, ``{PointAcc}: {Efficient}
  {Point} {Cloud} {Accelerator},'' in {\em {MICRO}-54: 54th {Annual}
  {IEEE}/{ACM} {International} {Symposium} on {Microarchitecture}},
  pp.~449--461, Oct. 2021.
\newblock arXiv:2110.07600 [cs].

\bibitem{elabd_graph_2022}
A.~Elabd, V.~Razavimaleki, S.-Y. Huang, J.~Duarte, M.~Atkinson, G.~DeZoort,
  P.~Elmer, S.~Hauck, J.-X. Hu, S.-C. Hsu, B.-C. Lai, M.~Neubauer, I.~Ojalvo,
  S.~Thais, and M.~Trahms, ``Graph {Neural} {Networks} for {Charged} {Particle}
  {Tracking} on {FPGAs},'' {\em Frontiers in Big Data}, vol.~5, p.~828666, Mar.
  2022.
\newblock arXiv:2112.02048 [hep-ex, physics:physics, stat].

\bibitem{li_deep_2020}
Y.~Li, L.~Ma, Z.~Zhong, F.~Liu, D.~Cao, J.~Li, and M.~A. Chapman, ``Deep
  {Learning} for {LiDAR} {Point} {Clouds} in {Autonomous} {Driving}: {A}
  {Review},'' May 2020.
\newblock arXiv:2005.09830 [cs].

\bibitem{elabd2022graph}
A.~Elabd, V.~Razavimaleki, S.-Y. Huang, J.~Duarte, M.~Atkinson, G.~DeZoort,
  P.~Elmer, S.~Hauck, J.-X. Hu, S.-C. Hsu, B.-C. Lai, M.~Neubauer, I.~Ojalvo,
  S.~Thais, and M.~Trahms, ``Graph neural networks for charged particle
  tracking on {{FPGAs}},'' {\em Frontiers in Big Data}, vol.~5, p.~828666, Mar.
  2022.

\bibitem{qu2020jet}
H.~Qu and L.~Gouskos, ``Jet tagging via particle clouds,'' {\em Physical Review
  D}, vol.~101, p.~056019, Mar. 2020.

\bibitem{auten2020hardware}
A.~Auten, M.~Tomei, and R.~Kumar, ``Hardware acceleration of graph neural
  networks,'' in {\em 2020 57th ACM/IEEE Design Automation Conference (DAC)},
  pp.~1--6, IEEE, 2020.

\bibitem{hygcn}
M.~Yan, L.~Deng, X.~Hu, L.~Liang, Y.~Feng, X.~Ye, Z.~Zhang, D.~Fan, and Y.~Xie,
  ``{{HyGCN}}: A {{GCN}} accelerator with hybrid architecture,'' in {\em 2020
  {{IEEE International Symposium}} on {{High Performance Computer
  Architecture}} ({{HPCA}})}, ({San Diego, CA, USA}), pp.~15--29, {IEEE}, Feb.
  2020.

\bibitem{engn}
S.~Liang, Y.~Wang, C.~Liu, L.~He, H.~Li, D.~Xu, and X.~Li, ``{{EnGN}}: A
  high-throughput and energy-efficient accelerator for large graph neural
  networks,'' {\em IEEE Transactions on Computers}, vol.~70, pp.~1511--1525,
  Sept. 2021.

\bibitem{awbgcn}
T.~Geng, A.~Li, R.~Shi, C.~Wu, T.~Wang, Y.~Li, P.~Haghi, A.~Tumeo, S.~Che,
  S.~Reinhardt, and M.~C. Herbordt, ``{{AWB-GCN}}: A graph convolutional
  network accelerator with runtime workload rebalancing,'' in {\em 2020 53rd
  {{Annual IEEE}}/{{ACM International Symposium}} on {{Microarchitecture}}
  ({{MICRO}})}, ({Athens, Greece}), pp.~922--936, {IEEE}, Oct. 2020.

\bibitem{zhang2021boostgcn}
B.~Zhang, R.~Kannan, and V.~Prasanna, ``{BoostGCN}: A framework for optimizing
  {GCN} inference on {FPGA},'' in {\em 2021 IEEE 29th Annual International
  Symposium on Field-Programmable Custom Computing Machines (FCCM)},
  pp.~29--39, IEEE, 2021.

\bibitem{igcn}
T.~Geng, C.~Wu, Y.~Zhang, C.~Tan, C.~Xie, H.~You, M.~Herbordt, Y.~Lin, and
  A.~Li, ``I-{{GCN}}: A graph convolutional network accelerator with runtime
  locality enhancement through islandization,'' in {\em {{MICRO-54}}: 54th
  {{Annual IEEE}}/{{ACM International Symposium}} on {{Microarchitecture}}},
  ({Virtual Event Greece}), pp.~1051--1063, {ACM}, Oct. 2021.

\bibitem{gcnax}
J.~Li, A.~Louri, A.~Karanth, and R.~Bunescu, ``{{GCNAX}}: A flexible and
  energy-efficient accelerator for graph convolutional neural networks,'' in
  {\em 2021 {{IEEE International Symposium}} on {{High-Performance Computer
  Architecture}} ({{HPCA}})}, ({Seoul, Korea (South)}), pp.~775--788, {IEEE},
  Feb. 2021.

\bibitem{rubik}
X.~Chen, Y.~Wang, X.~Xie, X.~Hu, A.~Basak, L.~Liang, M.~Yan, L.~Deng, Y.~Ding,
  Z.~Du, and Y.~Xie, ``Rubik: A hierarchical architecture for efficient graph
  neural network training,'' {\em IEEE Transactions on Computer-Aided Design of
  Integrated Circuits and Systems}, vol.~41, pp.~936--949, Apr. 2022.

\bibitem{graphact}
H.~Zeng and V.~Prasanna, ``{{GraphACT}}: Accelerating {{GCN}} training on
  {{CPU-FPGA}} heterogeneous platforms,'' in {\em Proceedings of the 2020
  {{ACM}}/{{SIGDA International Symposium}} on {{Field-Programmable Gate
  Arrays}}}, ({Seaside CA USA}), pp.~255--265, {ACM}, Feb. 2020.

\bibitem{kipf2016semi}
T.~N. Kipf and M.~Welling, ``Semi-supervised classification with graph
  convolutional networks,'' in {\em ICLR}, 2016.

\bibitem{hamilton2017inductive}
W.~L. Hamilton, R.~Ying, and J.~Leskovec, ``Inductive representation learning
  on large graphs,'' in {\em Proceedings of the 31st International Conference
  on Neural Information Processing Systems}, pp.~1025--1035, 2017.

\bibitem{tailor2021we}
S.~A. Tailor, F.~Opolka, P.~Lio, and N.~D. Lane, ``Do we need anisotropic graph
  neural networks?,'' in {\em International Conference on Learning
  Representations}, 2021.

\bibitem{xu2018powerful}
K.~Xu {\em et~al.}, ``How powerful are graph neural networks?,'' in {\em ICLR},
  2019.

\bibitem{corso2020principal}
G.~Corso {\em et~al.}, ``Principal neighbourhood aggregation for graph nets,''
  in {\em NeurIPS}, 2020.

\bibitem{deepburninggl}
S.~Liang, C.~Liu, Y.~Wang, H.~Li, and X.~Li, ``{DeepBurning}-{GL}: an
  {Automated} {Framework} for {Generating} {Graph} {Neural} {Network}
  {Accelerators},'' in {\em 2020 {IEEE}/{ACM} {International} {Conference} {On}
  {Computer} {Aided} {Design} ({ICCAD})}, pp.~1--9, Nov. 2020.
\newblock ISSN: 1558-2434.

\bibitem{lin2022hp}
Y.-C. Lin, B.~Zhang, and V.~Prasanna, ``Hp-gnn: Generating high throughput gnn
  training implementation on cpu-fpga heterogeneous platform,'' in {\em
  Proceedings of the 2022 ACM/SIGDA International Symposium on
  Field-Programmable Gate Arrays}, pp.~123--133, 2022.

\bibitem{velickovic_2022}
P.~Veličković, ``Message passing all the way up,'' 2022.

\bibitem{abadal2021computing}
S.~Abadal, A.~Jain, R.~Guirado, J.~L\'{o}pez-Alonso, and E.~Alarc\'{o}n,
  ``Computing graph neural networks: A survey from algorithms to
  accelerators,'' {\em ACM Comput. Surv.}, vol.~54, oct 2021.

\bibitem{auten}
A.~Auten, M.~Tomei, and R.~Kumar, ``Hardware acceleration of graph neural
  networks,'' in {\em 2020 57th {{ACM}}/{{IEEE Design Automation Conference}}
  ({{DAC}})}, ({San Francisco, CA, USA}), pp.~1--6, {IEEE}, July 2020.

\bibitem{grip}
K.~Kiningham, C.~Re, and P.~Levis, ``{{GRIP}}: A graph neural network
  accelerator architecture,'' {\em arXiv:2007.13828 [cs]}, July 2020.

\bibitem{abi2022gengnn}
S.~Abi-Karam, Y.~He, R.~Sarkar, L.~Sathidevi, Z.~Qiao, and C.~Hao, ``Gengnn: A
  generic fpga framework for graph neural network acceleration,'' {\em arXiv
  preprint arXiv:2201.08475}, 2022.

\bibitem{flowgnn}
R.~Sarkar, S.~Abi-Karam, Y.~He, L.~Sathidevi, and C.~Hao, ``{FlowGNN}: {A}
  {Dataflow} {Architecture} for {Universal} {Graph} {Neural} {Network}
  {Inference} via {Multi}-{Queue} {Streaming},'' Apr. 2022.
\newblock arXiv:2204.13103 [cs].

\bibitem{VitisHLS}
Vitis, {\em Vitis High-Level Synthesis User Guide (UG1399)}, Accessed: 2021.
\newblock \url{https://docs.xilinx.com/r/en-US/ug1399-vitis-hls}.

\bibitem{var_welford}
B.~P. Welford, ``Note on a {Method} for {Calculating} {Corrected} {Sums} of
  {Squares} and {Products},'' {\em Technometrics}, vol.~4, no.~3, pp.~419--420,
  1962.
\newblock Publisher: [Taylor \& Francis, Ltd., American Statistical
  Association, American Society for Quality].

\bibitem{gelu}
D.~Hendrycks and K.~Gimpel, ``Gaussian {Error} {Linear} {Units} ({GELUs}),''
  July 2020.
\newblock arXiv:1606.08415 [cs].

\bibitem{Vitis}
Xilinx, {\em Xilinx Vitis unified software platform}, Accessed: 2021.
\newblock \url{https://www.xilinx.com/products/design-tools/vitis.html}.

\bibitem{velivckovic2017graph}
P.~Veli{\v{c}}kovi{\'c} {\em et~al.}, ``Graph attention networks,'' in {\em
  arXiv preprint arXiv:1710.10903}, 2017.

\end{thebibliography}
\bibliographystyle{ieeetr}

\end{document}